# Thermal Expansion in YbGaGe

Svilen Bobev, Darrick J. Williams, J. D. Thompson, J. L. Sarrao

*Los Alamos National Laboratory, Los Alamos, New Mexico 87545, U.S.A.*

**Abstract:**

Thermal expansion and magnetic susceptibility measurements as a function of temperature are reported for YbGaGe. Despite the fact that this material has been claimed to show zero thermal expansion over a wide temperature range, we observe thermal expansion typical of metals and Pauli paramagnetic behavior, which perhaps indicates strong sample dependence in this system.

**Corresponding Author:**
Dr. John L. Sarrao
MST-10: Condensed Matter and Thermal Physics
MS K764
Los Alamos National Laboratory
Los Alamos, NM 87545
Phone: (505) 667-4838
Fax: (505) 665-7652
e-mail: *sarrao@lanl.gov*

# Thermal Expansion in YbGaGe

## 1. Introduction

Discovering materials whose size does not change as a function of temperature is an extremely active field of study, both for technological applications and for fundamental understanding [1]. Quite recently, Salvador *et al*. [2] observed zero thermal expansion (ZTE) in YbGaGe and attributed this behavior to a previously unreported mechanism for ZTE - valence fluctuations. Our study of this material calls into question both the mechanism and the observation of ZTE, suggesting that the reported behavior depends subtly on sample stoichiometry or preparation conditions.

## 2. Experimental

Polycrystalline YbGaGe was synthesized by direct fusion of pure elements in 1:1:1 ratio using the synthetic procedure described by Salvador *et al*. [2]. Unit cell parameters were obtained from *in situ* neutron-diffraction experiments (time-of-flight data, collected on the HIPPO and NPDF powder diffractometers at the Los Alamos Neutron Science Center) over the temperature range 20 – 770 K, [3] and by magnetic susceptibility measurements carried out using Quantum Design SQUID magnetometer between 1.7 and 350 K. Different reaction batches were measured upon cooling and heating to ensure reproducible results. Additionally, to explore carefully the possibilities of sample dependence and/or traces of unrecognized impurity phases, YbGaGe was prepared in a sealed tantalum tube under the same temperature conditions. The results obtained from all samples of YbGaGe were consistent, yet in obvious disagreement with those reported in Ref. 2.

## 3. Results and Discussion

Rietveld refinements [4] of the variable-temperature neutron-diffraction data allowed for the precise determination of the unit cell parameters and volume. A typical diffraction pattern is shown in Fig. 1. As shown in Fig. 2, the temperature-dependent cell volume and both *a*- and *c*- cell constants of the hexagonal YbGaGe increase monotonically with the temperature, and in the range $125 \leq T \leq 773$ K, the volume expansion coefficient ß = $4.0(1) \times 10^{-5}$ K$^{-1}$. Such behavior is typical of metals and intermetallic compounds.

Similar temperature dependence was observed for the Ta-tube-prepared sample of YbGaGe, although the absolute values of the lattice constants were slightly different. This is consistent with a certain phase width, [2] i.e., YbGa$_{1+x}$Ge$_{1-x}$, and suggests that small deviations from the ideal stoichiometry may play a decisive role in determining the properties. In that sense, the properties of bulk samples can depend in a very subtle way on the composition, cooling rates, and/or the presence of impurity phase(s). Specifically, our powder neutron-diffraction studies (Fig. 1) provided evidence for batch-dependent concentrations of mixed-valent Yb$_3$Ge$_5$ [5] as a secondary phase (note that the only difference in the synthesis of both is the use of sealed Ta-tube in one in place of an open alumina crucible in the other). Very recent work by Drymiotis *et al.* [6], and by Janssen *et al.* [7] on the crystal growth, magnetism, specific heat and the relative-elastic constant of YbGaGe (or rather YbGa$_{1+x}$Ge$_{1-x}$) lead to similar conclusions regarding sample variability.

Magnetic susceptibility data for our samples (shown in Fig. 3) also differ qualitatively from those reported in Ref. 2. Instead of increasing strongly with decreasing temperature, our data are weakly temperature dependent and consistent with the sum of two contributions, a dominant T-independent Pauli-like term from YbGaGe and another from a few *at*.% $Yb_3Ge_5$ that was found in the neutron-diffraction data, and whose magnetic susceptibility has a temperature dependence [5] very close to that shown in the upper-inset in Fig. 1.

**4. Conclusions**

The ZTE of YbGaGe observed by Salvador *et al*. [2] was determined from single crystal X-ray data at 100, 200, and 300 K, while other property measurements were performed on polycrystalline samples. Thus, it might be possible that the single crystal chosen for detailed study had the right composition, i.e. $YbGa_{1+x}Ge_{1-x}$ with a specific and yet to be determined value of *x*, that indeed displays ZTE, whereas the whole reaction product consists of a number of crystallites with small variations in composition and/or a small amount of unrecognized impurity. Our bulk samples display neither ZTE nor a strongly temperature dependent magnetic susceptibility and further establish that YbGaGe exists as $YbGa_{1+x}Ge_{1-x}$, and that ZTE, <u>if it occurs</u>, may be found only for some specific, yet unknown value of *x*.

*Note added as a proof.* While the preparation of this manuscript was in the final stage, work by Margadonna *et al*. that contain temperature-dependent $YbGa_{1.05}Ge_{0.95}$ data was published [8]. The X-ray diffraction data presented therein and the TOF neutron diffraction data we report herein are in excellent agreement over the temperature range 20 – 770 K. This work also provides an indication that the behavior of $YbGa_{1+x}Ge_{1-x}$ is very sensitive to slight stoichiometry deviations.

**Acknowledgements**

Work at LANL was performed under the auspices of the U.S. Department of Energy. Svilen Bobev thanks the Laboratory Directed Research and Development program (LANL/LDRD), and the Institute for Complex Adaptive Matter (I.C.A.M.) for financial support through Postdoctoral Fellowships. The authors are indebted to Dr. T. Proffen (LANL-LANSCE) for his help with the data collection on NPDF [3].

**Figure captions**

**Fig. 1.** TOF (time-of-flight) neutron powder diffraction pattern at $T = 20$ K of YbGaGe prepared as described in the text. Data were collected for approximately 2 grams of finely ground sample, loaded in vanadium can. Diffraction data at each temperature were acquired for 36000 pulses (LANSCE spallation source [3] is pulsed at 20 Hz and 100 µA power).

**Fig. 2.** Temperature dependence of the unit cell volume of YbGaGe (20 – 770 K). Error-bars represent standard deviations within $5\sigma$. The inset shows the dependence of both *a*- (red) and *c*-axis (blue) with temperature.

**Fig. 3.** Magnetic susceptibility (blue) and inverse magnetic susceptibility (red) of polycrystalline YbGaGe measured in a magnetic field of 1000 Gauss using a SQUID magnetometer (1.7 – 350 K).

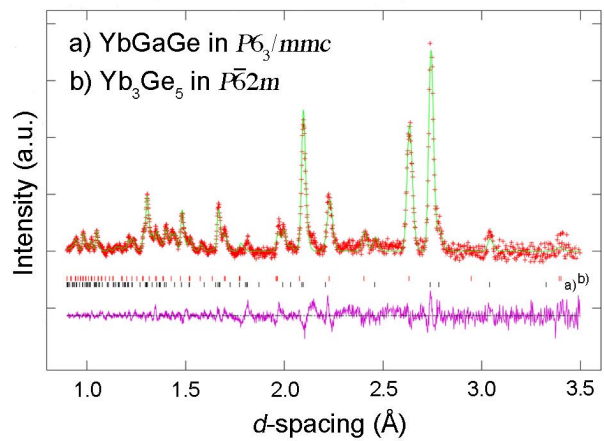

Fig. 1.

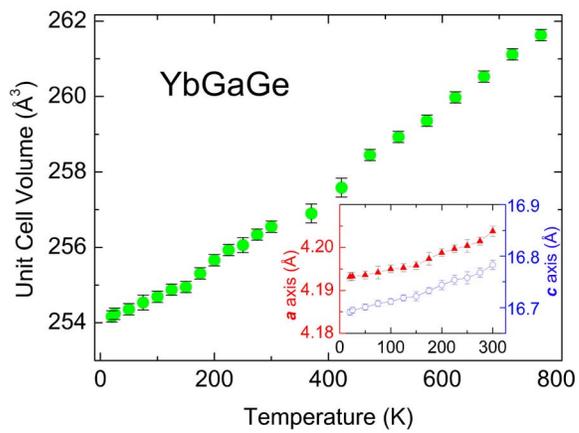

Fig. 2.

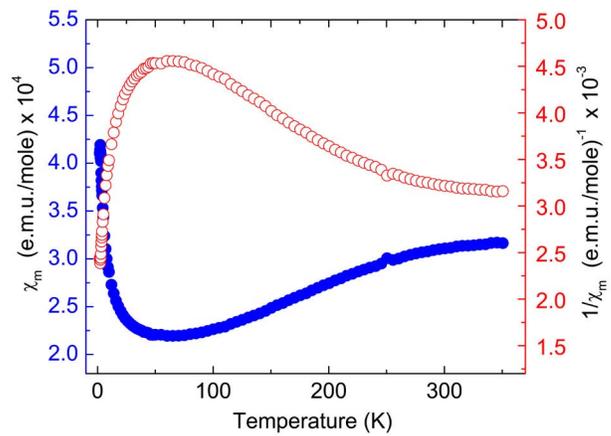

Fig. 3.

7